\documentclass[aip,jcp,%
 reprint,onecolumn,a4paper,10pt,longbibliography,%
 amssymb, amsmath%
]{revtex4-1}

\usepackage{color}
\usepackage{url}

\begin{document}

\title{On the plausible nature of the size effect in heterogeneous catalysis \newline
on gold nanoparticles}
\author{Kamil Moldosanov}
\email{altair1964@yandex.ru}
\affiliation{Kyrgyz-Russian Slavic University, 44 Kiyevskaya St., Bishkek 720000, Kyrgyzstan}
\author{Andrei Postnikov}
\email{andrei.postnikov@univ-lorraine.fr}
\affiliation{Universit\'e de Lorraine, Institut Jean Barriol, LCP-A2MC,
1 Bd Arago, F-57078 Metz, France}

\begin{abstract}%
We suggest that the size effect in heterogeneous catalysis on gold nanoparticles (GNP) of 
$\sim$1.3 -- 9~nm size can be related to spontaneous terahertz (THz) emission of GNPs
in the process of channeling the energy of longitudinal phonons therein 
into THz electromagnetic radiation. Since rotational and vibrational energy levels of molecules
adsorbed on the GNP surface are in the THz range, the irradiation by self-induced THz photons
would engage the molecules' swing and twisting, which in its turn would weaken
or break the intramolecular bonds, facilitating their entering a chemical reaction on the GNP surface.
\end{abstract}

\maketitle

The pioneering works by Haruta \emph{et al.} (see Refs.~\onlinecite{ChemRec3-75,CatalToday36-153}
for a review), who discovered catalysis by nanoparticles of gold,
hitherto considered to be a neutral metal, triggered attentive studies of gold and other noble metals'
nanoparticles. A vast bibliography exists on size effect in heterogeneous catalysis
at gold nanoparticles (GNP) -- see, e.g., 
Refs.~\onlinecite{JPhysChemC120-9174,AccChemRes47-834,Nanoscale6-532,%
Nanoscale6-13476,AccChemRes46-1712,JAmChemSoc135-16833,JAmChemSoc132-138,TopicCatal44-15,%
ChemPhysLett429-528,JAmChemSoc132-138,JCatal241-56,AngewChemInternEd43-5812,JPhysChemB108-15782,JCatal223-232}
and references therein. It seems that GNP of $\sim\,$1.3~nm to $\sim\,$9~nm 
diameter exhibit catalytic activity, especially pronounced at diameters $D$ below 5~nm, which decreases
as the GNP size increases, and disappears at $D\,{\gtrsim}\,$10~nm. In ongoing discussion about
possible nature of such behavior, we bring to the reader's attention 
a hypothesis not previously considered, to the best of our knowledge.

We suggest that the catalytic effect takes its origin in longitudinal acoustic (LA) phonons
hosted by a GNP, whose energies fall within the major peak of the density of
modes in the phonon spectrum of gold, centered around $\sim\,4.2$~THz.
Under certain conditions, part of the energy of these phonons (whatever the mechanism
supplying them) can be channeled into emission of photons in the terahertz (THz) range. 
As this is the frequency domain typical for rotational and vibrational movement,
the chances are that the molecules adsorbed on the GNP will have
degrees of freedom in resonance with the photons emitted. Consequently,
the THz ``irradiation'' by a host GNP may swing 
and twist the molecules, eventually resulting in weakening or breaking the chemical bonds. 
The effect of molecular parts entering a chemical reaction in the vicinity of GNPs
is then recognized as heterogeneous catalysis. The mechanism enabling THz emission
by GNPs was already addressed in our previous works albeit in
a different context.\cite{BeilJNano7-983,Nanotechnology29-285704} 

A crucial role in the process involving LA phonons and resulting
in emission of THz photons is played by electrons at the Fermi level, which can be, first, 
excited over several steps of their discrete (by force of spatial confinement) energy spectrum,
and then undergo either a radiative or non-radiative relaxation. The details of
the electron excitation are without relevance for the following simplified discussion.
The general idea, however, that some energy from the phonon ``bath'' can be channelled
into THz emission seems to be compatible with recent experimental evidence that the peak
of the density of modes of LA phonons in GNPs slightly redistributes part of its intensity
to the low-frequency side, as compared to the corresponding peak in bulk gold,
see Fig.~5 of Ref.~\onlinecite{SciRep6-39164}.

One may wish to discriminate between surface phonons, which propagate over roughly
circular orbits on the GNP's surface, and bulk or ``wall-to-wall'' ones, i.e., along
the GNP's ``diameter''. These two species are expected 
to act differently in the process of electron excitation. We will see that the allowed energies
of surface phonons are relatively densely distributed, so that matching the energy steps
with the electron system can be easily found and offer a ``typical'' channel
of the electron excitation. A generally more sparsely quantized bulk
phonons would less likely find a match and be absorbed ``in whole'' by the electron system; 
however, two-phonon processes can not be excluded whereby a higher-frequency phonon 
shares its energy onto that of the Fermi electron and the lower-frequency phonon.

Let us elaborate on the conditions of how the particle size, of the order of 1.3 to 9~nm,
enters the picture. For simplicity -- and in absence of more realistic models -- we consider
a GNP to be spherical of diameter $D$. The crucial observation is that the quantisation
step for energies of bulk phonons ought to be substantially (by a factor of $\simeq\,\pi$) 
more sparse than for surface ones. Indeed, the quantisation step
for the momentum of a ``wall-to-wall'' phonon confined within the length $D$ will be 
$h/D$, and the corresponding energy step
${\Delta}E^{\leftrightarrow}=hv^{\ast}_{\rm L}/D$, where $v^{\ast}_{\rm L}$ 
is the velocity of sound in the range of momenta / energies of interest. In our case, 
of major interest are phonons which populate the LA peak in the density of modes at $\sim\,4.2$~THz;
their dispersion branch bends down towards the Brillouin zone boundary, and the
velocity of sound (group velocity) is reduced from the nominal one in gold,
$v_{\rm L}\,\simeq\,3.23\,{\cdot}\,10^5$~cm/s (in the linear dispersion regime 
in the vicinity of the Brillouin zone center) to $v^{\ast}_{\rm L}\,\simeq\,10^5$~cm/s
in the middle of the said peak.
We will define the full width at the half maximum (FWHM) of the LA peak as spreading
from $\simeq\,3.9$~THz (16.2~meV) to $\simeq\,4.6$~THz (19.0~meV) -- see, e.g.,
Fig.~2 in Ref.~\onlinecite{BeilJNano7-983} that summarizes experimental findings,
hence FWHM$\,\simeq\,2.8$~meV for the following.
For surface phonons, the confinement length on the GNP surface
is of the order ${\pi}D$, hence the corresponding energy steps are more dense,
${\Delta}E^{\circlearrowright}=hv^{\ast}_{\rm L}/({\pi}D)$.

The energy matching conditions below have yet to be 
``softened'' to account for the Heisenberg uncertainty relation
following from the particles' small size, in the spirit of the discussion 
in Ref.~\onlinecite{Nanotechnology29-285704}, Appendix 1. With the momentum uncertainty given by
Eq.~(9) of this work, the uncertainty for the energy is
${\delta}E\,\simeq\,v^{\ast}_{\rm L}\,h/(2{\pi}D)$.

The interplay of different processes in the GNP may look, qualitatively, as follows. 
A typical bulk phonon would hardly excite the Fermi electron, since the phonon's energy
would be too high, or too much off-tune, with respect 
to the electron's discrete levels separation (given by the spatial
confinement and the electron density in terms of the Kubo formula -- 
see discussion in Ref.~\onlinecite{FANEM2015}). However, a ``circular'' phonon might
more easily satisfy a matching condition (for energy and momentum) 
to excite a Fermi electron across several steps in its energy spectrum.
Releasing this energy back in the course of relaxation via releasing a bulk
phonon will typically be difficult, since the energy steps of the latter
are sparse. However, the relaxation via emitting a photon might be possible,
as long as energy matching (a priori no problem) and the momentum conservation
(possible, taking into account the uncertainty relation) are secured. 
The maximum of emission is expected at energies of the most populated ``primary''
phonons, i.e., within the FWHM centered at $\sim\,4.2$~THz.

Otherwise, a Fermi electron may get excited by interacting with two bulk phonons,
absorbing the one and releasing the other, as long as the conservation conditions 
hold -- see discussion in Ref.~\onlinecite{Nanotechnology29-285704}.
Again, the relaxation channel via releasing a bulk phonon might be
complicated by force of the same argumentation as above, to the advantage
of relaxation via emitting a THz photon.

Somehow naively, the condition to prevent an easy relaxation into releasing 
a bulk phonon would be that ${\Delta}E^{\leftrightarrow}$
be substantially larger than the FWHM of the phonons' main peak,
adding to it the ``tolerance'' due to the uncertainty relation, ${\delta}E$,
hence
$$
\frac{hv^{\ast}_{\rm L}}{D} > \mbox{FWHM}+\frac{hv^{\ast}_{\rm L}}{2{\pi}D}\,; 
\quad\quad
D {\,\lesssim\,} 0.84\,\frac{hv^{\ast}_{\rm L}}{\mbox{FWHM}}\,,
$$
or, with the numerical estimations given above, $D<1.2$~nm.
Note that this condition stands for the maximal catalytic activity of GNP,
when the relaxation channel into releasing a LA phonon is effectively excluded;
larger $D$ values would not completely block this channel but gradually increase its
probability, thus depriving the THz photon emission of its efficiency.

Numerical estimates for two ``representative'' GNP sizes yield the following.

For $D=1.3$~nm, ${\Delta}E\,\approx\,v^{\ast}_{\rm L}{\cdot}h/D = 3.2$~meV,
${\delta}E\,\simeq\,0.5$~meV, so that FWHM+${\delta}E\,\approx\,3.3$~meV.
Even if the sparseness of phonon energy levels, ${\Delta}E$, is \emph{not} 
markedly larger than FWHM broadened by the uncertainty relation, 
the discretisation of the phonon energy steps remains ``visible'' on the energy scale
given by FWHM; this amounts to certain selectivity of which excitation energies
of an electron can be diverted into releasing phonons and which not. The excitations
``forbidden'' for being released into phonons ought to be relaxed via emitting
a THz photon.

For $D=10$~nm, ${\Delta}E\,\approx\,0.4$~meV,
${\delta}E\,\simeq\,0.1$~meV, and FWHM+${\delta}E\,\approx\,2.9$~meV,
hence much larger than ${\Delta}E$.
This means that, in practice, the phonon spectrum is quasicontinuous, and any
electron excitation can be converted into phonon, leaving only marginal
probability for THz emission channel. This is consistent with the experimental
observation that the catalytic activity
of GNPs of $D=10$~nm size is much lower than that of GNPs with $D=1.3$~nm. 

An extrapolation from the suggested mechanism of catalysis 
is that an activation of chemical reaction can be directly promoted by THz radiation.
In fact, this view seems to find support in a work by LaRue \emph{et al.}\cite{PRL115-036103}
who selectively induced the CO oxidation on Ru by applying intense THz pulses.
A further study on this subject, especially in what concerns selectivity and sensitivity,
may promise an enormous economical significance,
because more than 80\% of the present industrial chemical processes
use catalysts (see Ref.~\onlinecite{Deutschmann_in_Ullmanns_1}, Sec.~1.3).
However we would not go that far as to presume that metal nanoparticles 
could be fully substituted by THz emitters of a different nature. Rather it seems that the efficiency
of catalytic process on GNPs depends on the interplay of size-dependent (e.g. those 
discussed above) and size-independent conditions. Among the latter one may single out
local electric fields due to compression waves on the GNP's surface. 
A recent work by Pingel \emph{et al.}\cite{NatureCommun9-2722} addresses
``the relationship between strain and catalytic function'' on the surface of nanoparticles.
It is possible that the electric fields created by strains and compression waves
stay subcritical for catalytic activity, whereas the THz radiation (hence size-dependent
effect) makes it critical.

We note in passing that an independent experimental evidence seem to manifest the
(different) role of LA phonons propagating on the surface of gold nanospheres, namely,
the size effect in hyperthermia.\cite{FANEM2015} The nanoparticles
with sizes $D\,\simeq\,5\,-\,10$~nm seem to be the most efficiently \emph{heated}
by irradiation at 13.56~MHz. The context discussed in the present contribution
covers the ``complementary'' case of GNP \emph{cooling} accompanying the emission
of THz photons.


\newpage
\section*{Bibliography}
\begin{thebibliography}{10}

\bibitem{ChemRec3-75}
M.~Haruta, ``When gold is not noble: Catalysis by nanoparticles,'' {\em The
  Chemical Record}, vol.~3, p.~75, Apr 2003.

\bibitem{CatalToday36-153}
M.~Haruta, ``Size- and support-dependency in the catalysis of gold,'' {\em
  Catalysis Today}, vol.~36, p.~153, Apr 1997.

\bibitem{JPhysChemC120-9174}
Q.~Yao, C.~Wang, H.~Wang, H.~Yan, and J.~Lu, ``Revisiting the {Au} particle
  size effect on {TiO$_2$}-coated {Au/TiO$_2$} catalysts in {CO} oxidation
  reaction,'' {\em J. Phys. Chem. C}, vol.~120, p.~9174, May 2016.

\bibitem{AccChemRes47-834}
M.~Boronat, A.~Leyva-P{\'e}rez, and A.~Corma, ``Theoretical and experimental
  insights into the origin of the catalytic activity of subnanometric gold
  clusters: Attempts to predict reactivity with clusters and nanoparticles of
  gold,'' {\em Acc. Chem. Res.}, vol.~47, p.~834, Mar 2014.

\bibitem{Nanoscale6-532}
V.~Petkov, Y.~Ren, S.~Shan, J.~Luo, and C.-J. Zhong, ``A distinct atomic
  structure--catalytic activity relationship in 3--10 nm supported {Au}
  particles,'' {\em Nanoscale}, vol.~6, p.~532, 2014.

\bibitem{Nanoscale6-13476}
G.~Vil{\'e} and J.~P{\'e}rez-Ram{\'{\i}}rez, ``Beyond the use of modifiers in
  selective alkyne hydrogenation: silver and gold nanocatalysts in flow mode
  for sustainable alkene production,'' {\em Nanoscale}, vol.~6, p.~13476, 2014.

\bibitem{AccChemRes46-1712}
C.~T. Campbell, ``The energetics of supported metal nanoparticles:
  Relationships to sintering rates and catalytic activity,'' {\em Acc. Chem.
  Res.}, vol.~46, p.~1712, Aug 2013.

\bibitem{JAmChemSoc135-16833}
W.~Zhu, R.~Michalsky, {\"O}.~Metin, H.~Lv, S.~Guo, C.~J. Wright, X.~Sun, A.~A.
  Peterson, and S.~Sun, ``Monodisperse {Au} nanoparticles for selective
  electrocatalytic reduction of {CO$_2$} to {CO},'' {\em J. Am. Chem. Soc.},
  vol.~135, p.~16833, Nov 2006.

\bibitem{JAmChemSoc132-138}
X.~Zhou, W.~Xu, G.~Liu, D.~Panda, and P.~Chen, ``Size-dependent catalytic
  activity and dynamics of gold nanoparticles at the single-molecule level,''
  {\em J. Am. Chem. Soc.}, vol.~132, p.~138, Jan 2010.

\bibitem{TopicCatal44-15}
T.~V.~W. Janssens, B.~S. Clausen, B.~Hvolb{\ae}k, H.~Falsig, C.~H. Christensen,
  T.~Bligaard, and J.~K. N{\o}rskov, ``Insights into the reactivity of
  supported {Au} nanoparticles: combining theory and experiments,'' {\em Topics
  in Catalysis}, vol.~44, p.~15, Jun 2007.

\bibitem{ChemPhysLett429-528}
H.~Tsunoyama, H.~Sakurai, and T.~Tsukuda, ``Size effect on the catalysis of
  gold clusters dispersed in water for aerobic oxidation of alcohol,'' {\em
  Chem. Phys. Lett.}, vol.~429, p.~528, Oct 2006.

\bibitem{JCatal241-56}
S.~H. Overbury, V.~Schwartz, D.~R. Mullins, W.~Yan, and S.~Dai, ``Evaluation of
  the {Au} size effect: {CO} oxidation catalyzed by {Au/TiO$_2$},'' {\em
  Journal of Catalysis}, vol.~241, p.~56, Jul 2006.

\bibitem{AngewChemInternEd43-5812}
M.~Comotti, C.~Della~{P}ina, R.~Matarrese, and M.~Rossi, ``The catalytic
  activity of {``}naked{''} gold particles,'' {\em Angewandte Chemie
  International Edition}, vol.~43, p.~5812, Nov 2004.

\bibitem{JPhysChemB108-15782}
V.~Schwartz, D.~R. Mullins, W.~Yan, B.~Chen, S.~Dai, and S.~H. Overbury,
  ``{XAS} study of {Au} supported on {TiO$_2$}: Influence of oxidation state
  and particle size on catalytic activity,'' {\em J. Phys. Chem. B}, vol.~108,
  p.~15782, Oct 2004.

\bibitem{JCatal223-232}
N.~Lopez, T.~V.~W. Janssens, B.~S. Clausen, Y.~Xu, M.~Mavrikakis, T.~Bligaard,
  and J.~K. N{\o}rskov, ``On the origin of the catalytic activity of gold
  nanoparticles for low-temperature {CO} oxidation,'' {\em Journal of
  Catalysis}, vol.~223, p.~232, Apr 2004.

\bibitem{BeilJNano7-983}
K.~Moldosanov and A.~Postnikov, ``A terahertz-vibration to terahertz-radiation
  converter based on gold nanoobjects: a feasibility study,'' {\em Beilstein
  Journal of Nanotechnology}, vol.~7, p.~983, 2016.

\bibitem{Nanotechnology29-285704}
A.~V. Postnikov and K.~A. Moldosanov, ``Suggested design of
  gold-nanoobjects-based terahertz radiation source for biomedical research,''
  {\em Nanotechnology}, vol.~29, p.~285704, Jul 2018.

\bibitem{SciRep6-39164}
R.~Carles, P.~Benzo, B.~P{\'e}cassou, and C.~Bonafos, ``Vibrational density of
  states and thermodynamics at the nanoscale: the {3D}-{2D} transition in gold
  nanostructures,'' {\em Scientific Reports}, vol.~6, p.~4027, Dec 2016.

\bibitem{FANEM2015}
A.~Postnikov and K.~Moldosanov, ``Phonon-assisted radiofrequency absorption by
  gold nanoparticles resulting in hyperthermia,'' in {\em Fundamental and
  Applied Nano-Electromagnetics} (A.~Maffucci and S.~A. Maksimenko, eds.), The
  {NATO} Science for Peace and Security Programme, Series B: Physics and
  Biophysics, pp.~171 -- 201, Dordrecht, The Netherlands: Springer, 2016.
\newblock Proceedings of the NATO Advanced Research Workshop on Fundamental and
  Applied Electromagnetics, Minsk, Belarus, 25-27 May, 2015.

\bibitem{PRL115-036103}
J.~L. {LaR}ue, T.~Katayama, A.~Lindenberg, A.~S. Fisher, H.~{\"O}str{\"o}m,
  A.~Nilsson, and H.~Ogasawara, ``Thz-pulse-induced selective catalytic {CO}
  oxidation on {Ru},'' {\em Phys. Rev. Lett.}, vol.~115, p.~036103, Jul 2015.

\bibitem{Deutschmann_in_Ullmanns_1}
O.~Deutschmann, H.~Kn{\"o}zinger, K.~Kochloefl, and T.~Turek, ``Heterogeneous
  catalysis and solid catalysts, 1. {F}undamentals,'' in {\em Ullmann's
  Encyclopedia of Industrial Chemistry}, vol.~17, p.~457, Wiley-VCH Verlag GmbH
  \& Co. {KGaA}, Weinheim, 2011.

\bibitem{NatureCommun9-2722}
T.~N. Pingel, M.~J{\o}rgensen, A.~B. Yankovich, H.~Gr{\"o}nbeck, and E.~Olsson,
  ``Influence of atomic site-specific strain on catalytic activity of supported
  nanoparticles,'' {\em Nature Communications}, vol.~9, p.~2722, Jul 2018.

\end{thebibliography}

\end{document}